\documentclass{jfm}
\usepackage{graphicx}
\usepackage{epstopdf, epsfig}

\usepackage{tikz}
\usetikzlibrary{plotmarks}

\def\red{\textcolor{red}}
\definecolor{purple}{rgb}{0.6, 0.4, 0.8}
\def\purple{\textcolor{purple}}
\definecolor{persiangreen}{rgb}{0.0, 0.65, 0.58}

\definecolor{aurometalsaurus}{rgb}{0.43, 0.5, 0.5}

\definecolor{radicalred}{rgb}{1.0, 0.21, 0.37}
\def\rose{\textcolor{radicalred}}

\usepackage{gensymb}
\usepackage{graphicx}
\usepackage{dcolumn}
\usepackage{bm}

\usepackage{subcaption}
\usepackage{subfloat}
\usepackage{color}
\usepackage{bm}
\usepackage{ulem}
\usepackage{mhchem}

\shortauthor{A. Pouplard and P. A. Tsai}

\title{Viscous Fingering Instability of Complex Fluids in a Tapered Geometry}

\author{Alban Pouplard \and Peichun Amy Tsai \corresp{\email{peichun.amy.tsai@ualberta.ca}}}

\affiliation{Department of Mechanical Engineering, University of Alberta,\\ Edmonton, Alberta, Canada T6G 2G8}

\begin{document}

\maketitle

\begin{abstract}
Viscous fingering (VF) is an interfacial instability that occurs in a narrow confinement or porous medium when a less-viscous fluid pushes a more viscous one, producing finger-like patterns. Controlling the VF instability is essential to enhance the efficiency of various technological applications. However, the control of VF instability has been challenging and so far focused on simple Newtonian fluids of constant viscosity. Here, we extend to complex yield-stress fluids and examine the controlling feasibility by carrying out a linear stability analysis using a radial cell with a converging gap gradient. We avoid making the major assumption of a small Bingham number, $Bn \ll 1$, i.e., a negligible ratio of the yield to shear stress, and instead provide a new stability criterion predicting apparent complex VF.
This criterion depends on not only the complex fluid’s rheology ($\mu$), interfacial tension ($\gamma$), and contact angle to the wetting wall ($\theta_c$), but also the gap gradient ($\alpha$), the radius, gap-thickness, and velocity at the fluid-fluid interface ($r_0$, $h_0$ and $U_0$, respectively). Finally, we compare this theoretical criterion to our experimental data with nitrogen pushing a complex yield-stress fluid in a taper and find good agreement.
\end{abstract}

\section{Introduction
\label{Introduction}}

In many natural and industrial applications such as chromatographic separation \citep{Fernandez1996}, coating flows~\citep{Greener1980, 
Lee2005}, printing devices \citep{Pitts1961}, oil-well cementing,  
groundwater pollution \citep{Maes2010}, enhanced oil recovery (EOR) \citep{Green2018}, and \ce{CO_2} sequestration \citep{Huppert2014}, the displacement of a more viscous fluid by another less viscous one in a porous medium is essential. However, having an unfavorable viscosity or mobility contrast triggers an interfacial instability characterized by fingering patterns during the processes of fluid-fluid displacement.  This Saffman--Taylor instability \citep{Saffman1958}, or so-called viscous fingering (VF), is a major limiting factor for achieving maximum efficiency for some of the aforementioned applications. Consequently, the VF instability has been extensively studied for decades, particularly using Newtonian fluids in a paradigm Hele--Shaw cell consisting of two parallel plates separated by a fixed gap \citep{Paterson1981, Paterson1985, Saffman1986, Chen1987, Homsy1987}.

Several vital aspects of the fingering instability have been investigated using simple Newtonian fluids of constant viscosity.  Some focused on the influence of inertia to increase the finger-width above a critical Weber number \citep{Chevalier2006}. Others, for example, studied the effects of a surface tension that depends on the curvature on the possible stabilization (destabilization) of conventionally unstable (stable) situations of viscous fingers \citep{Rocha2013}. Furthermore, recent studies of viscous fingering have been extended to complex fluids, where wider fingers than those with Newtonian fluids have been obtained \citep{Park1994, Bonn1995}. Notably, more complex patterns of side-branching with multiple small fingers forming on the side of the major ones have been observed with yield-stress fluids \citep{Coussot1999}.

To enhance the efficiency of various technological applications, the feasibility of controlling or inhibiting VF has been investigated with simple Newtonian fluids. Several strategies have recently been developed to suppress the primary Saffman-Taylor instability, e.g., using time-dependent injection flow rate \citep{Cardoso1995, Dias2012, Zheng2015}, an elastic membrane \citep{Pihler2012, Pihler2013}, and a converging cell of linearly-varying gap-thickness \citep[][]{Housseiny2012, Stone2013, Bongrand2018}. However, such primary control has not been reported for complex fluids. Here, we demonstrate the control of VF instability for complex, yield stress fluids using a radial tapered cell. Firstly, we summarize our experimental results of nitrogen pushing a yield-stress, shear-thinning fluid. Secondly, we theoretically derive a stability criterion for two complex, yield-stress fluids pushing one another using linear stability analysis. Finally, we perform a systematic comparison between our experimental and theoretical results. Using the perturbation's growth rate at the most unstable mode, we found good agreement between our experimental and improved theoretical results without the small $Bn$-number assumption of $Bn \ll 1$.

\section{Experimental \label{Experimental}}

\begin{figure*}
    \begin{center}
    \includegraphics[width=5in]{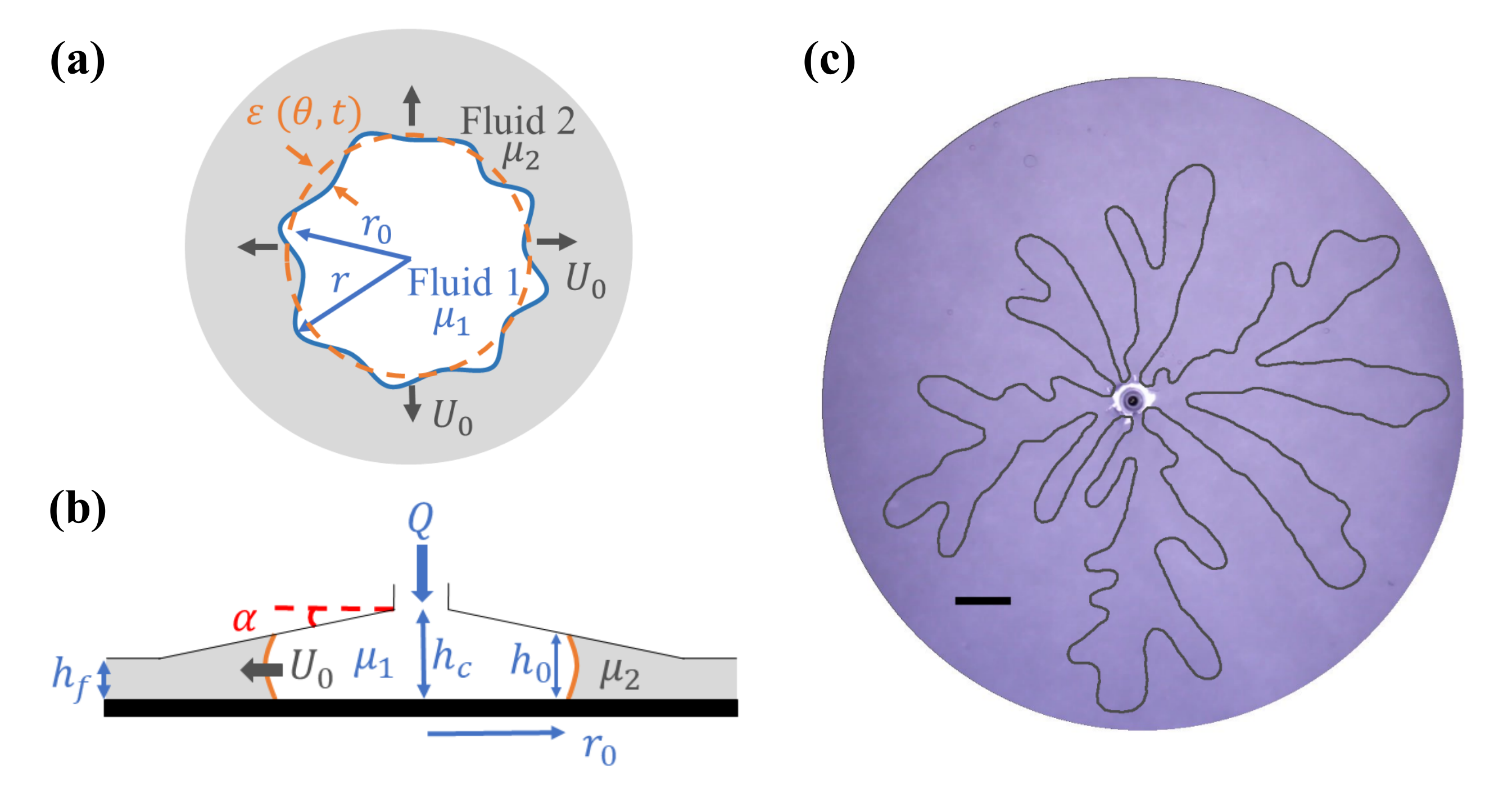}
    \end{center}
    \caption{(a)$-$(b) Schematics of the top-view and side-view fluid-fluid displacement process, where a more-viscous complex fluid of viscosity ($\mu_2$) is pushed by another immiscible one of lower viscosity ($\mu_1$). (c) An experimental snapshot of complex viscous fingering pattern with sided-branches obtained when the complex yield-stress (PAA) solution (S1) is invaded by nitrogen gas injected with a constant flow rate, $Q = 0.03$~slpm, in a uniform Hele$-$Shaw cell. The scale bar corresponds to $20$~mm.
    \label{Fig1}}
\end{figure*}

The experiments are conducted using both flat Hele$-$Shaw and converging tapered cells under a radial injection (See Fig.~\ref{Fig1}a-b). We use two aqueous solutions of PolyAcrylic Acid (PAA) of different concentrations, denoted as (S1) and (S2), for the wetting and receding fluid to examine the complex fluids' VF. These two yield-stress solutions are shear-thinning and have different viscosity values, but both can be modeled well by the Herschel$-$Bulkley (HB) model \citep{Herschel1926} of shear stress ($\tau$) varying with shear rate ($\dot{\gamma}$), via
$\tau = \tau_{c} + \kappa \dot{\gamma}^{n}$, where $\tau_{c}$, $\kappa$ and $n$ are the yield stress, consistency index, and power-law index, respectively.  The displacing fluid is gaseous nitrogen of viscosity $\mu_1 = 1.76 \times 10^{-5}$~Pa$\cdot$s (at $20~\degree$C), achieving the mobility contrast of ${\cal M} = \mu_2/\mu_1$ in the ranges of $2.68\times 10^{4}-1.16\times 10^{7}$ and $1.61\times 10^{3}-1.67\times 10^{5}$ for (S1) and (S2), respectively. Other vital control parameters include the cell's gap gradient ($\alpha$) and the constant flow-injection rate ($Q$), ranging from $0.02$ to $2$~slpm. The experimental setup, procedures, and HB fittings can be found in our recent work \citep{Pouplard_Tsai_2022}.

\begin{figure*}
    \begin{center}
    \includegraphics[width=4.5in]{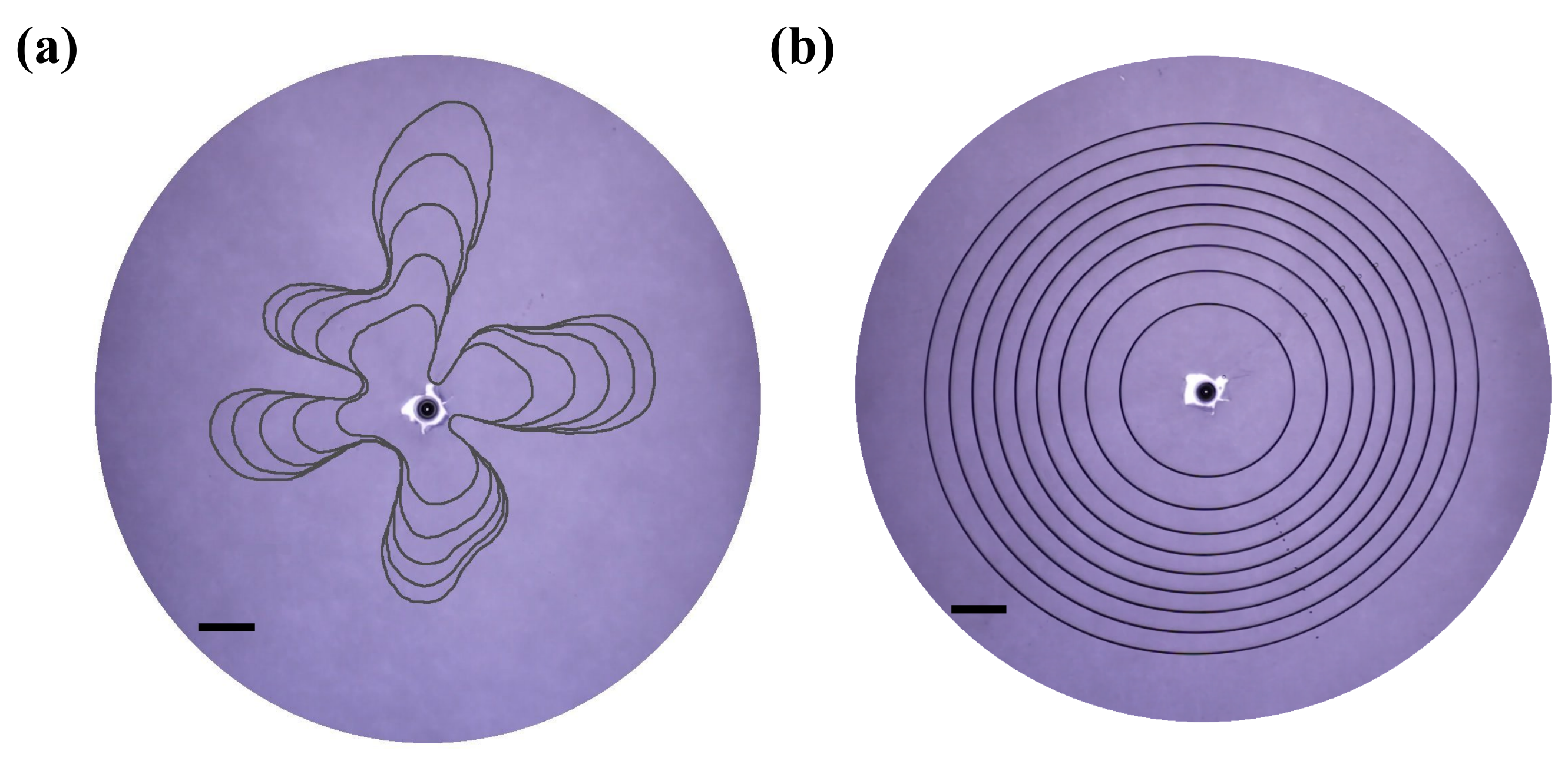}
    \end{center}
    \caption{{\bf Experimental Data of interfacial profiles} obtained when a gas displaces a yield-stress fluid (S2) using different cell geometries. (a) Overlay of experimental snapshots of a viscous fingering pattern observed in a flat Hele-Shaw cell with a uniform gap-thickness $h_c = 0.5$~mm and a constant injection rate $Q = 0.025$~slpm (with the time interval of $\delta t = 2$~s). (b) By contrast, overlay of experimental stable interfaces in a radially-tapered cell with a linearly converging gap-thickness: $h = h_c + \alpha r$, where $\alpha = -7.18 \times 10^{-2}$, $h_c = 10.39$~mm, and $Q = 0.025$~slpm (with $\delta t = 40$~s). Both scale bars correspond to $20$~mm.
    \label{Fig2}}
\end{figure*}

With flat Hele$-$Shaw cells, we observe side-branched fingering patterns (see Fig.~\ref{Fig1}c), as the typical characteristic for yield-stress VF observed in previous studies \citep{Jirsaraei2005, Eslami2017}, as well as more usual VF patterns (Fig.~\ref{Fig2}a). The latter type of VF pattern has been observed with Newtonian fluids in earlier works. By contrast, using converging cells we observe the elimination of side-branching fingers for both fluids (S1) and (S2), replaced by typical smoother classical VF, as also found recently by \citet{Eslami2020_1}. More remarkably, we are able to experimentally observe the inhibition of the primary VF instability under suitable rheological and flow parameters, e.g., $Q$, $\alpha$, $h_c$, as illustrated by Fig.~\ref{Fig2}b. The experimental stability diagram of controlling VF under various $\alpha$ and $Q$ for (S2) can be found in \citep[][]{Pouplard_Tsai_2022}.

The significant observations of our recent experiments using complex fluids include, first, the interface can become stable at lower $Q$ while keeping $\alpha$ and $h_c$ constant. Second, the transition between stable and unstable displacements occurs at higher $Q$ as $\alpha$ becomes steeper. Finally, the controlling criterion of the VF instability with yield-stress fluids is rather complex in that $\alpha$, $Q$, and the rheological parameters (via $\kappa$, $n$, and local $\dot{\gamma}$) all have a crucial influence on stabilizing the fluid-fluid interface.

\section{Theoretical Analysis \label{Theory}}

We carry out a linear stability analysis to shed light on the feasibility of controlling VF instability for complex fluids using a taper. The problem considered is one complex yield-stress fluid (Fluid 1) of low viscosity ($\mu_1$) pushing another one (Fluid 2) of high viscosity ($\mu_2$) in a radially-tapered cell, depicted in Fig.~\ref{Fig1}. With a constant gap gradient ($\alpha$), we consider a lubrication flow in the thin gap, whose height ($h$) varies linearly in $r$ direction as $h (r)= h_c + \alpha r$, where $h_c$ is the gap thickness at the cell center.

In the theoretical derivation, we use the effective Darcy's law replacing the conventional Newtonian fluids' constant viscosity with the effective shear-dependent viscosity, $\mu_\text{eff}$, for complex fluids.  This approach has been used to model non-Newtonian flow in a homogeneous porous medium \citep{Bonn1995,Amar1999,Lindner2000_1}, but here we extend the approach to a tapered geometry. Neglecting the fluids' elastic properties
\citep{Coussot1999}, the governing equations of the problem are 2D-depth-average Darcy's law and continuity Eq. considering the gap-thickness variation:
\begin{equation} \label{eq.1.appxD}
{\bf U}_j = -\frac{h^2}{12\mu_{\text{eff}j}} \vec{\nabla} P_j~,~~
\nabla \cdot \left(h {\bf U}_j \right) = 0. 
\end{equation}
${\bf U}_j(r,\theta) = (u_{rj} , u_{\theta j})$ and $P_j(r,\theta)$ are the depth-average velocity and pressure fields of the fluid indexed $j$, respectively. $j$ represents the two complex fluids during the fluid-fluid displacement process; $j = 1$ (2) denotes the pushing (displaced) complex fluid.

The complex fluid's viscosity ($\mu_{\text{eff} j}$) is modeled using the Herschel$-$Bulkley law for yield-stress fluids, with the local shear rate $\dot{\gamma} =  \frac{u_{rj}}{h}$, and expressed as 
\begin{equation} \label{eq.3.appxD}
\mu_{\text{eff} j} = \frac{\tau_{cj}}{\overset{.}{\gamma}} + \kappa_j {\overset{.}{\gamma}}^{n_j-1},
\end{equation}
where $\tau_{cj}$, $\kappa_j$ and $n_j$ correspond to the yield stress, consistency index, and power-law index, respectively.
Defining the Bingham number as the ratio of the yield to viscous stress: $Bn_j= \frac{\tau_{cj}}{\kappa_j \left(\frac{u_{rj}}{h} \right)^{n_j}}$, we express $\frac{\tau_{cj}}{\frac{h}{12} \frac{\partial P_j}{\partial r}} = - \frac{1}{1+\frac{1}{Bn_j}}$. Assuming a small ratio of gap-thickness change, i.e., $\frac{\alpha \left(r-r_0 \right)}{h_0} \ll 1$, one can linearize the expression of the gap thickness, $h = h_0 \left(1 +\frac{\alpha \left(r- r_0 \right)}{h_{0}} \right)$, neglecting the high-order terms of $\mathcal{O} (\alpha^2)$. With Eqs.~\eqref{eq.1.appxD}$-$\eqref{eq.3.appxD}, the continuity Eq. can be expressed using the pressure field ($P_j$): 
\begin{align}\label{eq.7.appxD}
& \frac{\partial^2 P_j}{\partial r^2} + \frac{n_j}{r} \frac{\partial P_j}{\partial r} + \frac{\left(2 n_j +1 \right) \alpha}{h_0} \frac{\partial P_j}{\partial r} + \frac{12 n_j \alpha \tau_{cj}}{h_0^2} + \frac{12 n_j \tau_{cj}}{h_0 r} + \frac{12 n_j \tau_{cj} \alpha}{{h_0}^2} \frac{r_0}{r} \nonumber\\
&  + \frac{n_j}{r^2}
\frac{\partial^2 P_j}{\partial \theta^2} \left(1 - \frac{1}{1+\frac{1}{Bn}} \right) + \frac{1}{r^2} \frac{\partial P_j}{\partial \theta} \left[\left(1-n_j \right) \frac{\frac{\partial^2 P_j}{\partial \theta \partial r}}{\frac{\partial P_j}{\partial r}} - n_j \frac{\tau_{cj} \frac{\partial^2 P_j}{\partial \theta \partial r}}{\frac{h}{12} \left(\frac{\partial P_j}{\partial r} \right)^2} \right] = 0.
\end{align}
For simple Newtonian fluids with $n_j = 1$ and $\tau_{cj} = 0$, we find the equation: $\frac{\partial^2 P_j}{\partial r^2} + \frac{1}{r} \frac{\partial P_j}{\partial r} + \frac{3 \alpha}{h} \frac{\partial P_j}{\partial r} + \frac{1}{r^2} \frac{\partial^2 P_j}{\partial \theta^2} = 0$, identical and recovered to the case for Newtonian fluids \citep{Stone2013} .

To solve for $P_j$, we decompose the solutions into the base and perturbed states using
\begin{equation} \label{eq.8.appxD}
P_j (r,\theta,t) = f_j (r) + g_{kj} (r) \epsilon (\theta,t),
\end{equation}
with the perturbation $\epsilon (\theta,t) = \epsilon_0 r_0(t) \exp{ \left(i k \theta+ \sigma t \right)}$.
$f_j(r)$ represents the base-state, pressure field independent of $\theta$.
The term of $g_{kj}(r) \epsilon$ represents the perturbation propagating from the interface with wavenumber ($k$) and growth rate ($\sigma$).
Focusing on the moment when the perturbation starts to propagates, 
the perturbation is still small $\left(\epsilon \ll 1 \right)$ so that $g'_{kj}(r) \epsilon \ll f'_j(r)$. Linearizing around the base state, we can express $r = r_0 \left(1 + \epsilon_0 z \right)$ with $\epsilon_0 \ll 1$. Substituting the pressure expression Eq.~\eqref{eq.8.appxD} into Eq.~\eqref{eq.7.appxD}, neglecting higher-order terms of $\mathcal{O} (\epsilon^2)$ but not $\mathcal{O} ({\epsilon_0}^2 k^2)$, and linearizing $\frac{1}{1+\epsilon_0 z}$ and $\frac{1}{\left(1+\epsilon_0 z \right)^2}$, the solutions of the base and perturbed states can be found via
\begin{equation} \label{eq.11.appxD}
\frac{\partial^2 f_j(z)}{\partial z^2} + n_j \epsilon_0 \frac{\partial f_j(z)}{\partial z} + \frac{\left(2 n_j +1 \right) \alpha}{h_0} \epsilon_0 r_0 \frac{\partial f_j(z)}{\partial z} =0.
\end{equation}
\begin{equation} \label{eq.12.appxD}
\frac{\partial^2 g_{kj}(z)}{\partial z^2} + n_j \epsilon_0 \frac{\partial g_{kj}(z)}{\partial z} + \frac{\left(2 n_j +1 \right) \alpha}{h_0} \epsilon_0 r_0 \frac{\partial g_{kj}(z)}{\partial z} - n_j k^2 {\epsilon_0}^2 g_{kj} \left(1+\frac{\tau_{cj}}{\frac{h}{12} \frac{\partial P_j}{\partial r}} \right) =0.
\end{equation}

From the Darcy's law, with the fact that $g'_{kj}(r) \epsilon \ll f'_j(r)$ at $r=r_0$, and $z = \frac{r-r_0}{\epsilon_0 r_0}$, we solve Eqn. \eqref{eq.11.appxD} for the base-state pressure solution, with the interface velocity $U_0$:
\begin{align} \label{eq.16.appxD}
f_j(r) = F_j \exp{\left( - \left(n_j + \frac{\left(2 n_j +1 \right) \alpha r_0}{h_0} \right) \frac{r - r_0}{r_0} \right)},~
F_j = \frac{12  \left[\tau_{cj} + \kappa_j \left(\frac{U_0}{h_0} \right)^{n_j} \right] r_0}{h_0 \left( n_j + \frac{\left(2 n_j +1 \right)\alpha r_0}{h_0} \right)}.
\end{align}

Assuming $\epsilon_0 \ll 1$, we have $\frac{\partial P_j}{\partial r} \approx f'_j(r)$, linearizing 
$h_0 + \alpha \left(r - r_0 \right) = h_0  \left( 1 + \frac{\alpha r_0 \epsilon_0 z}{h_0} \right)$ 
around $h_0$, and neglecting the terms $\mathcal{O} \left( \epsilon^2 \right)$, $\mathcal{O} ({\epsilon_0}^3 k^2)$ but not $\mathcal{O} \left( {\epsilon_0}^2 k^2 \right)$, Eq.~\eqref{eq.12.appxD} transforms into:
\begin{align} \label{eq.17.appxD}
& \frac{\partial^2  g_{jk}(z)}{\partial z^2} + n_j \epsilon_0 \frac{\partial g_{kj}(z)}{\partial z} + \frac{\left(2 n_j +1 \right) \alpha}{h_0} \epsilon_0 r_0 \frac{\partial g_{kj}(z)}{\partial z} \nonumber \\
&  - n_j k^2 {\epsilon_0}^2 g_{kj} \left( 1 - \frac{\tau_{cj}}{  \tau_{cj} + \kappa_j \left(\frac{U_0}{h_0} \right)^{n_j}} \exp{\left(\left(n_j + \frac{\left( 2 n_j +1 \right) \alpha r_0}{h_0} \right) \epsilon_0 z \right)} \right) = 0.
\end{align}
We define the following constants to simplify the above expression:
\begin{align} \label{eq.18.appxD}
A_j = \left(n_j + \frac{\left( 2 n_j +1 \right) \alpha r_0}{h_0} \right),~~
B_j = n_j k^2,~~
C_j = \frac{\tau_{cj}}{ \left( \tau_{cj} + \kappa_j \left(\frac{U_0}{h_0} \right)^{n_j} \right)} = \frac{\tau_{cj}}{\tau_{tj}},
\end{align}
where $A_j$, $B_j$ and $C_j$ correspond to the characteristic length of the exponential term of the base-state pressure [see Eq.~\eqref{eq.16.appxD}], the impact of the perturbation's wavenumber, and the ratio of the yield stress ($\tau_c$) to the total stress ($\tau_t$), respectively.

The general solution ($u_\text{Sol j}$) for the above Eq.~\eqref{eq.17.appxD} is a linear combination of the first-order and second-order Bessel functions, $J(\beta, z)$ and $Y(\beta, z)$, respectively. To simplify the expressions for the Bessel functions, we define the following terms: 
\begin{align}
N_j = \frac{\sqrt{{A_j}^2 + 4 B_j}}{A_j},~~
M_j (z) = \frac{2 \sqrt{B_j C_j}}{A_j} \exp{\left(\frac{A_j \epsilon_0 z}{2} \right)}. \label{eq.29.appxD}
\end{align}
Note that $J\left(\beta, z \right)$ or $Y\left( \beta, z \right)$ are regular functions throughout the z-plane cut along the negative real axis \citep{Barton1965}, meaning $J (\beta, {\cal M}(z))$ or $Y\left( \beta, {\cal M}(z) \right)$ are only defined for ${\cal M}(z) > 0$. We can then simplify the general solution ($u_\text{Sol j}$) for Eq.~\eqref{eq.17.appxD} 
\begin{align}
  u_\text{Sol j} (z) &= 
  \sum_{g=1\rightarrow 2}Cg_{kj} \exp{\left(\frac{-A_j \epsilon_0 z}{2} \right)} \; J \left( (-1)^g N_j, \frac{2 \sqrt{\exp{\left(A_j \epsilon_0 z \right)} B_j C_j}}{\left|A_j\right|} \right) \nonumber \\ 
  & + \sum_{g= 3\rightarrow 4}Cg_{kj} \exp{\left(\frac{-A_j \epsilon_0 z}{2} \right)} \; Y \left((-1)^g N_j, \frac{2 \sqrt{\exp{\left(A_j \epsilon_0 z \right)} B_j C_j}}{\left|A_j\right|} \right).
 \end{align}

It is physically impossible for the perturbation to grow in space from its origin, implying that for a fluid indexed 1 pushing a fluid indexed 2 we have $u_\text{Sol 1}|_{r \rightarrow 0} = 0$,
$u_\text{Sol 2}|_{r \rightarrow + \infty} = 0$,~
$u_\text{Sol 1}|_{z \rightarrow -\infty} = 0$,~
$u_\text{Sol 2}|_{z \rightarrow + \infty} = 0$.
Performing analyses on the functional limits when $z \rightarrow \pm \infty$, we observe that only
${\exp{\left(\frac{-A_j \epsilon_0 z}{2} \right)} J \left(N_j, \frac{2 \sqrt{\exp{\left(A_j \epsilon_0 z \right)} B_j C_j}}{\left|A_j\right|}  \right) \rightarrow 0}$ as ${z \rightarrow - \infty}$, and only
${\exp{\left(\frac{-A_j \epsilon_0 z}{2} \right)} \; J \left(- N_j, \frac{2 \sqrt{\exp{\left(A_j \epsilon_0 z \right)} B_j C_j}}{\left|A_j\right|}  \right) \rightarrow 0}$ as ${z \rightarrow + \infty}$.
In brief, the final solution of $r$-dependent perturbed pressure, $g_{kj}(r)$, fulfilling Eq.~\eqref{eq.17.appxD} is expressed as
\begin{align} \label{eq.19.appxD}
 u_\text{Sol j} (z) &= C1_{kj} \exp{\left(\frac{-A_j \epsilon_0 z}{2} \right)}~J ( (-1)^{j+1} N_j,~| M_j(z) |).
\end{align}

Assuming that the length scale of the interfacial perturbation ($|\frac{r_0}{k}|$) is much smaller than that of the depth variation characterized by $|\frac{h_0}{\alpha}|$, i.e., $|\frac{\alpha r_0}{k h_0}| = O \left(\epsilon \right)$,
and transforming $z = \frac{r - r_0}{r_0 \epsilon_0}$, Eq. \eqref{eq.19.appxD} can be further simplified to
\begin{equation} \label{eq.20.appxD}
u_\text{sol j} (r) = C1_{kj} \exp{\left(\frac{-A_j}{2} \frac{r - r_0}{r_0} \right)} J \left( (-1)^{j+1} \frac{2 \sqrt{n_j} k}{A_j}, \frac{2 \sqrt{n_j} k \sqrt{\frac{\tau_{cj}}{\tau_{tj}}} \exp{\left(\frac{A_j}{2} \frac{r - r_0}{r_0} \right)}}{\left|A_j\right|} \right).
\end{equation}

The final pressure solution is then expressed as $P_j (r, \theta, t) = f_j (r) + u_\text{sol j} (r) \epsilon ( \theta, t)$. We linearize the pressure expression around the interface, $ r_{int} = r_0 + \epsilon$, neglect the higher-order terms of $\mathcal{O} (\epsilon^2)$, and obtain
\begin{equation} \label{eq.23.appxD}
P_j|_{r = r_0 + \epsilon}  = \frac{12 \tau_{tj} r_0}{A_j h_0} - \frac{12 \tau_{tj}}{h_0} \epsilon + C1_{kj} \; J \left( (-1)^{j+1} \frac{2 \sqrt{n_j} k}{A_j}, \frac{2 \sqrt{n_j} k \sqrt{\frac{\tau_{cj}}{\tau_{tj}}}}{\left|A_j\right|} \right) \epsilon.
\end{equation}
We further use the properties of the Bessel functions of the first ($J$) and second order ($Y$) \citep{Barton1965}: $J'(N_j, M_j(z)) = M_j' (z) \left( \frac{N_j}{M_j(z)} J(N_j, M_j(z)) - J(N_j + 1, M_j(z)) \right)$,
$\frac{2 N_j}{ M_j(z)} J (N_j,  M_j(z)) = J(N_j + 1,  M_j(z)) + J(N_j - 1, M_j(z))$. We then linearize the first derivative of $P_j$ at the interface, $r = r_{int} = r_0 + \epsilon$, neglecting $\mathcal{O} (\epsilon^2)$ terms. 
To obtain $C1_{k1}$ and $C1_{k2}$, we use the kinematic condition of the same fluid velocity at the interface:
\begin{align} \label{eq.25.appxD}
\frac{\partial r_{\text{int}}}{\partial t} = u_{r j}|_{r = r_0 + \epsilon},~~
\frac{\partial r_0}{\partial t} + \frac{\partial \epsilon}{\partial t} = u_{r j}|_{r = r_0 + \epsilon},~~
U_0 + \sigma \epsilon &= u_{r j}|_{r = r_0 + \epsilon}.
\end{align}
Assuming $\epsilon \ll 1$, neglecting the higher-order terms of $O(\epsilon^2)$, linearizing
$h|_{r= r_0 + \epsilon} = h_0 \left(1 + \frac{\alpha \epsilon}{h_0} \right)$ with $\frac{\alpha \epsilon}{h_0} \ll 1 $, and using the previous approximation $\left| \frac{k h_0}{\alpha r_0} \right| \gg 1$ to assume $\left(-1\right)^{j+1} \frac{2 \sqrt{n_j} k}{A_j} + 1 $ to be equal to $ \left(-1\right)^{j+1} \frac{2 \sqrt{n_j} k}{A_j}$, we obtain:
\newline
$ J \left( \left(-1\right)^{j+1} \frac{2 \sqrt{n_j} k}{A_j}, \frac{2 \sqrt{n_j} k \sqrt{\frac{\tau_{cj}}{\tau_{tj}}}}{|A_j|} \right) \approx J \left(\left(-1\right)^{j+1} \frac{2 \sqrt{n_j} k}{A_j} + 1, \frac{2 \sqrt{n_j} k \sqrt{\frac{\tau_{cj}}{\tau_{tj}}}}{|A_j|} \right)$.
Finally, the constant $C1_{kj}$ can be expressed as
\begin{equation} \label{eq.26.appxD}
C1_{kj} =~\frac{12 \kappa_j}{{h_0}^{n_j +1}} \frac{\left( n_j \sigma r_0 U_0^{n_j - 1} + 2 n_j {h_0}^{n_j} \frac{\tau_{cj}}{\kappa_j} \frac{\alpha r_0}{h_0} + n_j {h_0}^{n_j} \frac{\tau_{cj}}{\kappa_j} + n_j {U_0}^{n_j} + n_j {U_0}^{n_j}  \frac{\alpha r_0}{h_0} \right)}{\; J \left(\left(-1\right)^{j+1} \frac{2 \sqrt{n_j} k}{A_j}, \frac{2 \sqrt{n_j} k \sqrt{\frac{\tau_{cj}}{\tau_{tj}}}}{|A_j|} \right) \left( \frac{A_j}{2}
+\left(-1\right)^{j} \sqrt{n_j} k + \sqrt{n_j} k \sqrt{\frac{\tau_{cj}}{\tau_{tj}}} \frac{A_j}{|A_j|}\right)}.
\end{equation}


The Capillary pressure jump at the fluid-fluid interface is described by the Young-Laplace equation, which accounts for both the lateral curvature ($\Psi$) and the curvature due to the depth of the narrow cell $\left(\frac{1}{h} \right)$ \citep{Stone2013}. However, here we neglect the contributions of the viscous stresses to the pressure difference across the interface. 
Since ${r_0}^2 \gg \epsilon^2$, by linearizing around $r_0$ and neglecting $\mathcal{O}(\epsilon^2)$, the Young-Laplace Eq. at the interface for $r= r_0 + \epsilon (\theta,t)$ transforms into
\begin{equation}\label{eq.39.appxD}
P_1 - P_2 = \frac{2 \gamma \cos{\theta_c}}{h_0} + \frac{\gamma}{r_0} + \gamma \epsilon \left(\frac{k^2 -1}{{r_0}^2} - \frac{2 \alpha \cos{\theta_c}}{{h_0}^2} \right) +O \left(\epsilon^2 \right),
\end{equation}
where $\gamma$ is the interfacial tension, and $\theta_c$ is the contact angle of the wetting fluid to the side wall. The first two terms on the right hand side (RHS) correspond to the base state, i.e., the pressure difference at the interface when the interface is stable. The third term on the RHS is the additional Laplace pressure due to the perturbation at the interface.

We substitute the expression of the linearized pressure Eq.~\eqref{eq.23.appxD} into Eq.~\eqref{eq.39.appxD} and remove all the base-state components. Using Eq.~\eqref{eq.26.appxD}, we derive the dimensionless dispersion-relation with the dimensionless growth rate ($\overline{\sigma} = \frac{\sigma r_0}{U_0}$) and the dimensionless wavenumber ($\overline{k} = k$) of the perturbation for the complex, yield-stress fluids:
\begin{align} \label{eq.31.appxD}
& \overline{\sigma} = \biggl[ \frac{\gamma \overline{k}^2 {h_0}^2 }{12 {r_0}^2 U_0} - \frac{\gamma {h_0}^2}{12 {r_0}^2 U_0} - \frac{2 \gamma \alpha \cos\theta_c}{12 U_0} + \left( \mu_1|_{r=r_0} - \mu_2|_{r=r_0} \right) + \frac{n_2  \mu_2|_{r=r_0} + n_2 \frac{\alpha r_0}{U_0} \left( \tau_{t2} + \tau_{c2} \right)}{\frac{A_2}{2} + \sqrt{n_2} \overline{k} \left( 1 + \sqrt{\frac{\tau_{c2}}{\tau_{t2}}} \frac{A_2}{|A_2|} \right)} \nonumber \\
& + \frac{n_1 \mu_1|_{r=r_0} + n_1 \frac{\alpha r_0}{U_0} \left( \tau_{t1} + \tau_{c1} \right)}{ - \frac{A_1}{2} + \sqrt{n_1} \overline{k} \left( 1 - \sqrt{\frac{\tau_{c1}}{\tau_{1}}} \frac{A_1}{|A_1|} \right)} \biggl] \frac{-1}{\frac{n_1 \kappa_1 \left(\frac{U_0}{h_0} \right)^{n_1 - 1}}{- \frac{A_1}{2} + \sqrt{n_1} \overline{k} \left( 1 - \sqrt{\frac{\tau_{c1}}{\tau_{t1}}} \frac{A_1}{|A_1|} \right)} + \frac{n_2 \kappa_2 \left(\frac{U_0}{h_0} \right)^{n_2 - 1}} {\frac{A_2}{2} + \sqrt{n_2} \overline{k} \left( 1 + \sqrt{\frac{\tau_{c2}}{\tau_{t2}}} \frac{A_2}{|A_2|} \right)}}.
\end{align}

By setting $\frac{ \partial \overline{\sigma}}{\partial \overline{k}} = 0$ in the dispersion relation Eq.~\eqref{eq.31.appxD}, we can find the dimensionless wavenumber of maximum growth rate ($\bar{k}_{max}$). Furthermore, for each unstable $\bar{k}$, the corresponding $\bar{\sigma}$ is given by Eq.~\eqref{eq.31.appxD} and depends primarily on the rheological parameters and $\dot{\gamma}$ at the fluid-fluid interface. From the dispersion-relation Eq.~\eqref{eq.31.appxD}, the interface theoretically would always be stable for a negative growth rate, $\overline{\sigma} <0$.

\section{Comparison between the Experimental and Theoretical Results \label{Comparing}}

\begin{figure*}
    \centering
    \includegraphics[width=4in]{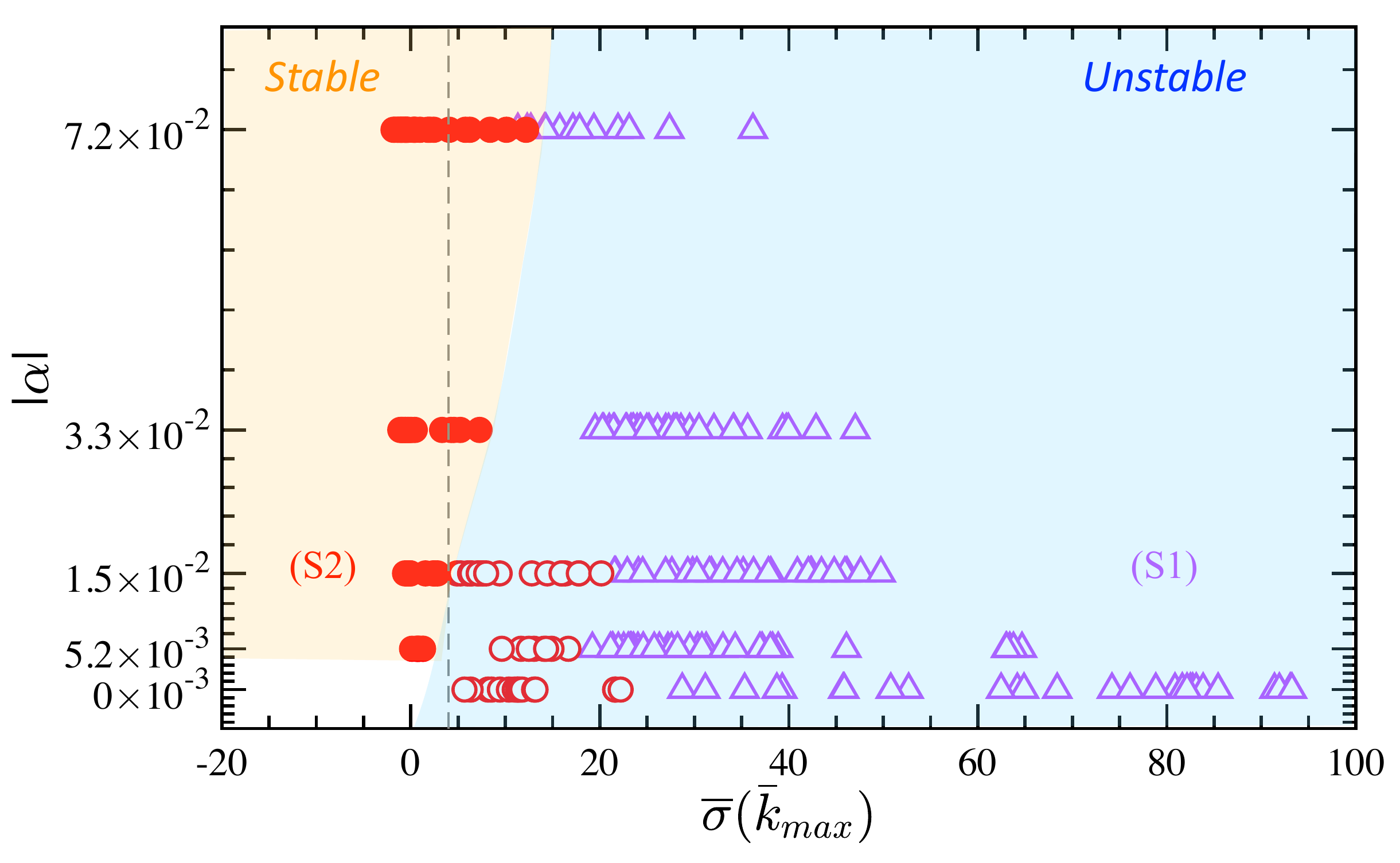}
    \caption{{\bf Comparison between experimental and theoretical results} of the perturbation's growth rate of the most unstable mode, $\overline{\sigma}(\overline{k}=\bar{k}_{max})$ [Eq.~\eqref{eq.31.appxD}]. The values of $U_0$, $r_0$ and $h_0$ are taken from the experiments by \citet{Pouplard_Tsai_2022}. The wavenumber corresponding to the wavenumber of maximum growth ($\overline{k}=\bar{k}_{max}$) is obtained numerically using Matlab.  We compare with the experimental results performed with various $|\alpha|$ and differentiate stable displacements (red filled circle, {\LARGE\red{$\bullet$}}), which are obtained solely during experiments with the less-viscous complex fluid (S2), and unstable wavy interfaces (purple open triangle, {\large\purple{$\vartriangle$}}) and (red open circle, {\LARGE\red{$\circ$}}) for the fluid (S1) and (S2), respectively.
    \label{Fig3}}
\end{figure*}

The linear stability analysis performed above gives a complex expression of the dimensionless perturbation's growth rate [Eq.\eqref{eq.31.appxD}] depending on the dimensionless wavenumber of the perturbation ($\overline{k}$), rheological properties of the fluids ($\kappa_j, n_{j}, \tau_{cj}$), the gap gradient ($\alpha$), as well as the velocity, radius, and gap thickness at the interface ($U_0, h_0, r_0$, respectively). Additional factors such as the wetting angle ($\theta_c$) and the interfacial tension ($\gamma$) also affect the VF stability. Whenever the growth rate is less than 0, the interface would remain stable for a decaying perturbation.

With the experimental values of $U_0$, $h_0$ and $r_0$, we use Matlab to determine numerically the wavenumber of maximum growth ($\bar{k}_{max}$) with an accuracy of $0.001$. We then substitute it into the dispersion Eq.~\eqref{eq.31.appxD} to obtain the growth rate at the most unstable mode, $\overline{\sigma}(\bar{k}_{max})$, and plot such results for various $\alpha$ in Fig.~\ref{Fig3}. In agreement, experimental results show stable interfaces (filled symbols) for complex fluids for negative and small values of $\overline{\sigma}(\bar{k}_{max})$ whereas unstable interfaces (open symbols) for greater positive values of $\overline{\sigma}(\bar{k}_{max})$. Although theoretically one would expect a transition between stable ({\LARGE\rose{$\bullet$}}) and unstable displacements ({\large \purple{$\vartriangle$}} and {\LARGE\rose{$\circ$}}) at $\overline{\sigma}(\bar{k}_{max}) = 0$, our experimental data are consistent with the theoretical predictions but the empirical transitional point is at $\overline{\sigma} (\bar{k}_{max}) \approx 4$, as the lowest boundary for unstable displacement, shown in Fig.~\ref{Fig3}. For lower values of $|\alpha|$ and up to $1.49 \times 10^{-2}$, this boundary can be used as a border to differentiate the stable and the unstable displacements.

From Fig.~\ref{Fig3}, however, one could notice that $\overline{\sigma} (\bar{k}_{max})$ is superior to 4 for stable displacements for higher values of $\alpha$. This difference may be explained by the assumptions made concerning $\alpha$. We assume a small ratio of gap change ($ \frac{\alpha (r-r_0)}{h_0} \ll 1$) as well as much larger characteristic length scale over which the depth varies than that of the perturbation ($\frac{k h_0}{\alpha r_0} \gg 1$). A similar deviation between theoretical and experimental results for higher values of $\alpha$ has already been noticed previously for simple fluids, when the theoretical stability criterion derived by \citet{Stone2013} has been used and compared to radial experimental results \citep{Bongrand2018}. Finally, the discrepancy observed by the experimental result of~$\overline{\sigma} (\bar{k}_{max}) \approx 4$ may also stem from the assumptions made as we neglect the influence of the gravity and the fluid's elastic properties in the analysis.

\section{Conclusions \label{Conclusion}}

We theoretically derive the stability criterion for the viscous fingering instability of complex yield-stress fluids using a narrow taper. Experimentally, for the very viscous fluid (S1) of a greater mobility contrast, ${\cal M}$ ranging from $2.68\times 10^{4}$ to $1.16\times 10^{7}$, we could observe the elimination of small side-branching fingers but not the primary finger with the tapered cells. However, stable interfaces with the inhibition of the primary VF instability can be achieved for the less-viscous one (S2) of ${\cal M}$ spanning $1.61\times 10^{3}-1.67\times 10^{5}$. The experimental observation of the stability diagram (under various $\alpha$ and $Q$) by \citet[][]{Pouplard_Tsai_2022} shows a clear transition between a complete and incomplete sweep.

Using a linear stability analysis with the effective Darcy's law, we obtain a stability criterion corresponding to the perturbation's growth rate of the most unstable mode. The stability criterion derived for the complex fluids depends on the following three types of important parameters: first, the complex fluid's rheological characteristics and property constants of $\kappa$, $\tau_c$ and $n$, $\gamma$, $\theta_c$; second, the gap gradient ($\alpha$); lastly, the interface velocity, position gap thickness and velocity ($r_0$, $h_0$ and $U_0$, respectively). We examine this theoretical criterion with the experiments done using the two yield-stress fluids of distinct {$\cal M$}. Taking the experimental values of $U_0$, $r_0$ and $h_0$, we calculate the perturbation's growth rate for the most unstable mode. From the different values obtained, we observe a transition between stable and unstable interfaces for $\overline{\sigma}(\bar{k}_{max}) = 4$, which slightly deviated from the theoretically expected transition around 0. This discrepancy may stem from the assumptions we made in the derivation. For instance, the impact of the gravity and the elastic properties of the fluids have been neglected. We also use a static contact angle for moving fluids and make few assumptions related to the gap gradient ($\alpha$). Notably, we assume a small ratio of gap change and the characteristic length scale over which the depth varies being much larger than the characteristic length scale of the perturbation. Both assumptions especially impact the results for larger values of $\alpha$.

Comparing with simple fluids of a constant viscosity, both dispersion-relation and stability criterion for complex yield-stress fluid are far more complex depending on additional parameters, such as rheological and flow measurements (e.g., $n$, $k$, $\tau_c$, $\tau_t$, $U_0$) at the interface, revealed in Eq.\eqref{eq.31.appxD}. Nevertheless, it is feasible and practical to control or inhibit complex fluids' viscous fingering instability using a tapered cell, thereby offering various strategies for manipulating fluid-fluid interfaces in microfluidics, narrow passages, packed beads, and porous media, while our theoretical framework present here offers fundamental insights through Eq.\eqref{eq.31.appxD} and $\overline{\sigma}(\bar{k}_{max})$, the corresponding growth-rate of the most unstable mode. 

\begin{acknowledgments}
{A.P.} and {P.A.T.} thankfully acknowledge the funding support from the Natural Sciences and Engineering Research Council of Canada (NSERC) Discovery grant (RGPIN-2020-05511). {P.A.T.} holds a Canada Research Chair in Fluids and Interfaces (CRC TIER2 233147). This research was undertaken, in part, thanks to funding from the Canada Research Chairs (CRC) Program.\\
\end{acknowledgments}

{\bf Declaration of Interests.} The authors report no conflict of interest.

\vspace{-0.2in}
\bibliographystyle{jfm}
\bibliography{Complex_VF.bib}

\end{document}